\documentclass[a4paper,final,authoryear]{elsarticle}
\pdfoutput=1
\input{src/latexdiff_preamble}

\usepackage[english]{babel}

\usepackage[DIV=13]{typearea}
\usepackage{microtype}

\usepackage{amsmath}
\usepackage{amsfonts}
\usepackage{amsthm}
\usepackage{booktabs}
\usepackage{stmaryrd}
\usepackage{multirow}
\usepackage{placeins}
\usepackage{enumitem}

\usepackage{graphicx}

\usepackage[colorlinks=true,hypertexnames=false]{hyperref}

\usepackage{hypernat}

\usepackage{cleveref}

\newcommand{\tens}[1]{{\boldsymbol{#1}}}

\renewcommand{\d}[0]{{\,\mathrm{d}}}
\renewcommand{\epsilon}{\varepsilon}

\newcommand{\hooke}[0]{{\tens{\mathcal{C}}}}

\newcommand{\Id}[0]{{\tens{I}}}
\newcommand{\body}[0]{{\mathcal{B}}}

\DeclareMathOperator\Tr{Tr}

\theoremstyle{plain}

\theoremstyle{definition}

\theoremstyle{remark}
\newtheorem{remark}{Remark}

\newcommand{\hsub}{_{\text{\scriptsize h}}}
\newcommand{\ysub}{_{\text{\scriptsize y}}}
\newcommand{\ic}{_\mathrm{\scriptsize i/c}}
\newcommand{\sigmay}{{\sigma\ysub}}
\newcommand{\I}{{_\mathrm{I}}}

\newcommand{\Rbb} {\mathbb{R}}

\newcommand{\Mcal}{\mathcal{M}}
\newcommand{\Ncal}{\mathcal{N}}

\newcommand{\shp}{\hspace*{-0.1em}+\hspace*{-0.1em}}

\graphicspath{{src/figs/}}

\begin{document}
\title{Crack Nucleation in the Adhesive Wear of an Elastic-Plastic
  Half-Space}
\author[epfl]{Lucas Fr\'erot\corref{cor}}
\ead{lfrerot1 <at> jhu <dot> edu}
\author[epfl]{Guillaume Anciaux}
\author[epfl]{Jean-François Molinari}
\address[epfl]{Institute of
  Civil Engineering, École Polytechnique Fédérale de Lausanne (EPFL),
  CH 1015 Lausanne, Switzerland}
\cortext[cor]{Corresponding Author}

\begin{abstract}
  The detachment of material in an adhesive wear process is driven by a fracture
  mechanism which is controlled by a critical length-scale. Previous efforts in
  multi-asperity wear modeling have applied this microscopic process to rough
  elastic contact. However, experimental data shows that the assumption of
  purely elastic deformation at rough contact interfaces is unrealistic, and
  that asperities in contact must deform plastically to accommodate the large
  contact stresses. We therefore investigate the consequences of plastic
  deformation on the macro-scale wear response using novel elastoplastic contact
  simulations. The crack nucleation process at a rough contact interface is
  analyzed in a comparative study with a classical $J_2$ plasticity approach and
  a saturation plasticity model. We show that plastic residual deformations in
  the $J_2$ model heighten the surface tensile stresses, leading to a higher
  crack nucleation likelihood for contacts. This effect is shown to be stronger
  when the material is more ductile. We also show that elastic interactions
  between contacts can increase the likelihood of individual contacts nucleating
  cracks, irrespective of the contact constitutive model. This is supported by a
  statistical approach we develop based on a Greenwood--Williamson model
  modified to take into account the elastic interactions between contacts and
  the shear strength of the contact junction.
\end{abstract}



\maketitle

Adhesive wear, as defined by \citet{burwell_survey_1957}, happens when two
asperities, belonging to different surfaces being rubbed together, form a
junction (which can be adhesive, chemical, mechanical, etc.) that supports the
shear caused by the relative sliding of the surface, such that cracks propagate
in the bulk of the materials, causing the detachment of a third-body. In this
setting, the removal of a wear debris particle from the surfaces is mainly
driven by a fracture process, and as such obeys a balance between the energy
release rate (i.e.\ the energy released by the crack front advancing) and the
fracture toughness (i.e.\ the energy required to create new surfaces). This
\citet{griffith_vi_1921} energy balance has been verified in atomistic
simulations
\citep{aghababaei_critical_2016,aghababaei_asperitylevel_2018,brink_adhesive_2019}
and reduced to a critical length-scale $d^* \propto G\Delta w / \tau_j^2$, with
$G$ being the shear modulus, $\Delta w$ the fracture energy and $\tau_j$ the
junction shear strength. This gives a simple geometric criterion for the
formation of hemispherical wear particles: if the contact diameter between two
hemispherical asperities is larger than $d^*$ then a wear particle detaches from
the surface upon shearing of the system. The issue of transposing asperity-scale
wear mechanisms to multi-asperity contact is key in the goal of formulating
predictive wear models~\citep{meng_wear_1995,vakis_modeling_2018}.
\Citet{popov_adhesive_2018} and \citet{pham-ba_adhesive_2019} have recently
proposed energy-based models for the formation of wear particles in
multi-asperity settings. The former investigates the formation of hemispherical
wear particles in an elastic rough surface contact by computing an
energy-favored particle diameter based on the elastic deformation energy of the
contact solution. The latter formulates the energetic competition between the
formation of a single vs.\ multiple wear particles (for 2D line contacts), thus
giving an energy approach to the crack shielding mechanism that leads to
\emph{disjoint but sufficiently close} contacts forming a single wear
particle~\citep{aghababaei_asperitylevel_2018}.

In a previous work~\citep{frerot_mechanistic_2018}, we have applied
the critical length-scale concept to rough elastic contact by defining
a critical cluster area $A^* \propto {(d^*)}^2$ above which
micro-contacts should form a wear particle. This work however suffers
from two model inadequacies: (a) the contact solution is given by an
elastic contact model, (b) it assumes that $A^*$ exists and is
proportional to the square of $d^*$. The latter is related to the
topography and shape of contacts. Contacts resulting from interfaces
with rough surfaces are not disk-shaped and the crack is not expected
to produce hemispherical wear particles. Moreover, this does not
account for disjoint contacts that may form a single
particle~\citep{aghababaei_asperitylevel_2018,pham-ba_adhesive_2019}.
Thence, it is unclear if the Griffith balance can be characterized
with a comparison as ``naive'' as
$A\, {\scriptstyle \overset{\scriptscriptstyle ?}{\geq}}\, A^*$ with
$A$ being the area of a single contact cluster.

The former shortcoming (a) provides to the wear models developed
in~\citep{frerot_mechanistic_2018} an unrealistic contact solution.
Since \emph{both} $A^*$ and the contact solution indirectly depend on
$\sigmay$\footnote{We have $A^* \propto {\sigmay}^{-4}$ if one assumes
  $\tau_j\propto\sigmay$ and the total contact area
  $A_c \propto \sigmay^{-1}$ according to saturation models}, the
outcome of an elastic contact problem (which is independent of
$\sigmay$) leads to a paradox: more ductile materials (with lower
$\sigmay$) have a higher $A^*$ and thus wear less than more brittle
materials (with higher $\sigmay$). This is due to the contact solution
being insensitive to changes in $\sigmay$, but also to the lack of
surface roughness evolution in sliding. When the contact of two
asperities does not create a wear particle, plastic smoothing of the
asperities occurs, thus creating larger contacts. In any case, a
contact model incorporating plastic effects is needed.

Experimental data clearly shows that some form of plasticity must
occur at rough contact
interfaces~\citep{bowden_area_1939,greenwood_contact_1966,weber_molecular_2018,zhang_experimental_2019}.
Modeling these interfaces with a non-linear constitutive behavior is
however a challenge because of the multi-scale nature of rough
surfaces. \Citet{pei_finite_2005} were the first to use the
finite-element method to study elastic-plastic rough contact with a
classical von~Mises formulation~\citep{simo_computational_1998}.
\Citet{jacq_development_2002} have developed a volume integral method
that we have refined with a Fourier approach to be able to handle the
large discretization requirements of multi-scale rough
surfaces~\citep{frerot_fourieraccelerated_2019}. The majority of
published works on elastic-plastic contact does not rely on classical
formulations of plastic flow, but rather on the concept of surface
flow pressure, which is associated to the hardness of a
material~\citep{archard_contact_1953,greenwood_contact_1966,majumdar_fractal_1991,persson_elastoplastic_2001}.
The surface flow pressure is usually taken as the maximum value of the
mean pressure caused by an indenter of a given shape (it is therefore
shape-dependent). \Citet{tabor_hardness_1951} has shown that for a
spherical indenter, the mean pressure saturates at a value close to
$3\sigmay$ (with $\sigmay$ being the yield stress). The models
previously mentioned are thereafter referred to as ``saturation
models'', in the sense that they apply this concept of a maximum
average pressure to a multi-asperity contact model~\cite[see
e.g][chap. 9]{tabor_hardness_1951} and assume that a given contact
cannot have a pressure exceeding the saturation pressure
noted~$p_m$\footnote{More often than not, the saturation pressure is
  referred to as ``hardness''. As \citet{burwell_empirical_1952}
  discuss, the saturation (or flow) pressure cannot be absolutely
  known but is of the same order of magnitude as the value given by
  usual hardness tests. We therefore keep separate notations for
  clarity.}. They have been used in conjunction with boundary integral
approaches~\citep{almqvist_dry_2007} to study
friction~\citep{weber_molecular_2018}, but have to our knowledge never
been compared to classical plasticity formulations, and the relevance
of the choice between the two plasticity models has never been
studied.

\Citet{akchurin_stresscriterionbased_2016} and
\citet{li_mesoscale_2019} have used a saturation plasticity model to
compute the contact solution and applied a stress based criterion for
the removal of debris particles: from the contact pressure profile,
they computed the resulting von~Mises stress caused in a \emph{purely
  elastic} medium. Then the zones of the material where the von~Mises
stress exceeds the yield stress are removed, changing the surface
profile. This has the advantage of foregoing any geometrical
consideration, at the expense of providing an ad-hoc removal process
that is not derived from the fracture energy balance, as well as using
a stress distribution that does not account for plastic deformations.

In this work, we wish to investigate the multi-asperity wear process from a
fracture-mechanics perspective and understand the influence of plasticity in the
contact model on the global wear response. To this end, we focus on the crack
nucleation process in the contact of a rigid self-affine rough surface with an
elastic-plastic flat half-space (without surface energy effects). The contact is
resolved using full-order simulations~\citep{frerot_fourieraccelerated_2019}.
One measure of particular importance is the crack spatial density. While it is
not a measure of wear itself, crack nucleation is a necessary process of wear,
and understanding what are the roles of the normal load, the critical nucleation
stress, the junction resistance, and plastic behavior in crack nucleation is a
fundamental step towards predictive wear models. We first highlight the
importance of the choice of a plasticity model and the implications it may have
on the contact response (\cref{sec:ep}). We then show how the crack nucleation
density in a rough surface elastic-plastic contact depends on the fracture
mechanics properties of the material, as well as the applied load and the
junction shear strength (\cref{sec:crack_rough}). To rationalize the differences
between the elastic, the saturation and the von~Mises plasticity approaches, we
study the contact behavior of a single asperity to understand under which
conditions a crack can nucleate and what is the influence of residual plastic
deformations on this process (\cref{sec:single}). These findings are applied to
a simple multi-asperity contact model~\citep{greenwood_contact_1966} in order to
obtain qualitative analytical predictions of the scaling of the crack density
with respect to system parameters like the applied normal load
(\cref{sec:multi}). These predictions are confronted to the elastic and
elastoplastic simulations results which are able to reproduce the contact
shielding effect under shear loading, as seen in molecular dynamics
simulations~\citep{aghababaei_asperitylevel_2018,pham-ba_adhesive_2019}. The
elastic-plastic results show that ductile materials in contact with rough
surfaces produce more crack nucleation sites than brittle materials due to the
residual stresses caused by plastic deformations. This effect is not captured by
the elastic contact model nor the saturation plasticity model, indicating that
the resolution of the aforementioned wear paradox should include the full
plastic contact response. This further implies that the true contact area is not
the only key quantity in wear modeling.


\section{Elastic-plastic contact}\label{sec:ep}
At our disposal are (at least) two formulations of the elastic-plastic
contact of solids, the choice of which may have an impact on the
subsequent results we wish to obtain. The first formulation, which has
been used in the finite-element studies of \citet{pei_finite_2005},
follows the classical modeling hypothesis of metal
plasticity~\citep{simo_computational_1998}, which have both
experimental~\citep{bui_etude_1969} and theoretical
backgrounds~\citep{reddy_internal_1994}, and additionally are valid in
other context than contact. The second, developed by
\citet{bowden_area_1939} and extended by \citet{almqvist_dry_2007} in
conjunction with a boundary integral approach, postulates that the
surface contact pressure should nowhere exceed a maximum value $p_m$.
This is based on observations that for spherical indentation the mean
contact pressure does not exceed a value around
$3\sigmay$~\citep{tabor_hardness_1951}. Recent finite-element
simulations~\citep{krithivasan_analysis_2007,song_elastic_2013,ghaednia_review_2017}
show that $p_m / \sigmay$ may depend on the ratio $\sigmay / E^*$
(with $E^* := E / (1 - \nu^2)$ being the contact modulus) as well as
the wavenumber in the case of sinusoidal contact surfaces. Despite
these reports, saturation models are often used in computational
tribology~\citep{weber_molecular_2018,akchurin_stresscriterionbased_2016,li_mesoscale_2019}
due to their simplicity and ease of implementation. Besides increasing
the magnitude of the true contact area compared to elastic contact,
plasticity influences other aspects of the contact interface (such as
contact pressures). These additional aspects may be key ingredients in
wear modeling. For this reason we wish to provide a comprehensive
comparison between the von~Mises associated plasticity and the
saturation plasticity with $p_m = 3\sigmay$ in a rough contact
situation, and determine the consequences of the choice of one model
over the other. We start by giving the full mechanical formulation for
both models, then proceed to the comparison.

\paragraph{Definitions} In this work, we consider a deformable
three-dimensional solid $\body$ spanning a half-space, with its (flat)
boundary noted $\partial\body$. Moreover, we suppose a horizontal
periodicity in the cell $\body_p = {[0, L]}^2\times \Rbb^+$. We note
$\tens \sigma$ the Cauchy stress tensor, which is related to the
small-strain tensor $\tens \epsilon$ and the plastic strain tensor
$\tens \epsilon_p$ by the relation
$\tens \sigma = \hooke:(\tens \epsilon - \tens \epsilon_p)$ where
$\hooke$ is the usual isotropic linear elasticity tensor. The strain
tensor is given by kinematic compatibility as a function of the
displacement field: $\tens \epsilon = \nabla^\mathrm{sym}\tens
u$. Finally, $\tens\sigma$ is expected to be divergence-free to
satisfy conservation of momentum without volume forces.

We additionally define some surface quantities: $\tens t$ and
$p := \tens t\cdot \tens e_3$ are respectively tractions and normal
pressures applied on $\partial \body$. Other surface quantities are
noted with an over-bar $\overline \bullet$ when not explicitly defined
on $\partial\body$, e.g.\ $\overline{\tens u}$ is the surface
displacement.

\subsection*{Saturation: perfect plasticity}
The simplest form of saturation model, conceptually close to the
notion of ``perfect plasticity'', is given
as~\citep{almqvist_dry_2007}
\begin{subequations}
  \begin{equation}\label{eq:min_principle}
    \min_p \left\{ \frac{1}{2}\int_{\partial\body_p}{p \overline\Mcal[p]\d{S}}
      - \int_{\partial\body_p}{p h\d{S}} \right\},
  \end{equation}
  which is a problem of finding the surface pressures $p$ minimizing
  the complementary energy of the system under the constraints
  \begin{align}
    p &\geq 0,\\
    p &\leq p_m,\\
    \int_{\partial\body_p}{p\d{S}} &= W.
  \end{align}
\end{subequations}
The linear operator $\overline\Mcal$ gives the normal surface
displacement due to the applied pressure $p$ if $\body$ is assumed
elastic; $h$ is a continuous function representing the rough surface
brought in contact with $\partial\body$ and $W$ is the total applied
normal load in the periodic cell boundary $\partial\body_p$. The gap
is defined as $g := \overline\Mcal[p] - h$ and should satisfy weak
Hertz--Signorini--Moreau conditions~\citep{weber_molecular_2018}:
\begin{subequations}
  \begin{align}
    g & \geq 0\quad\text{where } p < p_m,\\
    p & \geq 0,\\
    p \, g & = 0\quad\text{where } p < p_m.
  \end{align}
\end{subequations}
The solution to the above constrained optimization problem yields a
negative gap where $p = p_m$. The magnitude of the negative gap is
often assumed to be the magnitude of the residual plastic
displacements. Since the weak optimality conditions do not represent a
physical system (the gap should be non-negative everywhere to avoid
body interpenetration), it is necessary to replace $h$ in
\cref{eq:min_principle} by $h_\mathrm{mod}^s := h + h_\mathrm{pl}$,
with $h_\mathrm{pl}$ being in principle a correction due to residual
plastic displacements, therefore
$h_\mathrm{pl} := -(\overline\Mcal[p] - h)$ where $p = p_m$.
\citet{weber_molecular_2018} propose an iterative scheme to solve for
$h_\mathrm{pl}$ which we have implemented and made available in the
open-source contact library \emph{Tamaas}~\citep{frerot_tamaas_2019}
(\url{https://c4science.ch/tag/tamaas/}).

The ``perfect plasticity'' aspect of the model comes from the fact
that $p_m$ is homogeneous on $\partial\body$ and constant.
\citet{weber_molecular_2018} have amended this hypothesis to include a
form of hardening. The saturation pressure is simply expressed as a
linear function of $h_\mathrm{pl}$ (i.e.\ the initial saturation
stress is zero, and rises in proportion with $h_\mathrm{pl}$). We will
however not discuss this particular model here.

\subsection*{$J_2$ von Mises plasticity}
For the Cauchy stress tensor $\tens \sigma$, the von Mises yield
function $f\ysub$ is defined as
\begin{equation}
  f\ysub(\tens \sigma) = \sqrt{\frac{3}{2}}||\tens s||,\text{ where } \tens s := \tens \sigma - \frac{1}{3}\Tr(\tens \sigma)\Id.
\end{equation}
The equivalent cumulated plastic strain is expressed as the integral
of the plastic strain rate $\dot{\tens \epsilon}^p$ from some
reference time $t_0$:
\begin{equation}
  e^p := \sqrt{\frac{2}{3}}\int_{t_0}^t{||\dot{\tens \epsilon}^p||\d{t}}.
\end{equation}
The admissibility and consistency conditions are written as:
\begin{subequations}
  \begin{align}
    f\ysub(\tens \sigma) - f\hsub(e^p) & \leq 0,\\
    \big(f\ysub(\tens \sigma) - f\hsub(e^p)\big)\dot e^p & = 0,
  \end{align}
\end{subequations}
where $f\hsub$ is the hardening function. In this work, we will only
consider functions of the form $f\hsub(e^p) = \sigmay + E_h e^p$, with
$\sigmay$ the initial yield stress and $E\hsub$ the hardening
modulus\footnote{This corresponds to linear isotropic hardening.}. The
associated flow rule that determines $\dot{\tens \epsilon}^p$ is given
by~\citep{johnson_contact_1985a}:
\begin{equation}
  \dot{\tens \epsilon}^p = \frac{3 \dot e^p}{2 f\ysub(\tens \sigma)}\tens s(\tens \sigma).
\end{equation}
The numerical integration of the relations above is typically done
with a backwards Euler scheme and is classical to the solid mechanics
literature~\citep{simo_computational_1998}. Its coupling with the
equilibrium and contact conditions is however non-trivial.

\paragraph{Solution strategy}\Citet{jacq_development_2002}
established a numerical method for the solution of the elastic-plastic
rough contact problem, which we summarize here. The method consists in
solving the contact and the plasticity problems separately. The
contact problem is solved for fixed plastic deformations:~it is
effectively an elastic contact problem with a rough surface
$h_\mathrm{mod} := h - \overline u_3^p$, with $\overline u_3^p$ the
vertical component of the actual\footnote{In this approach the
  residual displacement is directly computed from $\tens \epsilon^p$,
  whereas in the saturation plasticity model it is merely
  \emph{assumed} equal to the negative gap.} plastic residual
displacement. Various solution strategies for the elastic rough
contact problem are available in the
literature~\citep{bemporad_optimization_2015}, and we use here the
modified conjugate gradient algorithm of
\citet{polonsky_numerical_1999} coupled with the spectral approach of
\citet{stanley_fftbased_1997} for the gradient computation involving
the operator $\overline \Mcal$.

The plastic problem is solved with fixed boundary tractions, meaning
that the contact area does not evolve during the resolution of the
plastic strain increment. The procedure we employ, fully detailed
in~\citep{frerot_fourieraccelerated_2019}, relies on an implicit
incremental volume integral equation formulation proposed by
\citet{telles_implicit_1991}. The total strain increment is shown to
be expressed as:
\begin{equation}\label{eq:implicit_plasticity}
  \Delta \tens{\epsilon} = \nabla^\mathrm{sym}\Mcal[\Delta \tens{\tens t}] + \nabla^\mathrm{sym}\Ncal[\hooke:\Delta\tens\epsilon^p(\Delta\tens\epsilon; S)],
\end{equation}
where $S := (e^p, \tens \epsilon^p)$ is the current plastic state,
$\Delta\tens t$ is the increment of surface tractions (in our case
$\Delta \tens t = \Delta p \,\tens e_3$ as we are in a normal contact
situation). The function
$\Delta\tens\epsilon^p(\Delta\tens\epsilon; S)$ represents the
radial-return algorithm classically used in incremental plastic
analysis~\citep{simo_computational_1998}. \Cref{eq:implicit_plasticity}
is a non-linear equation that can be solved with the DF-SANE
algorithm~\citep{lacruz_spectral_2006} which has the advantage of
being jacobian-free.

The operators $\Mcal$ and $\Ncal$, which are at the heart of the
method developed in~\citep{frerot_fourieraccelerated_2019}, are linear
integral operators which compute in $\body$ the displacement due to
periodic distributions of surface traction and volume eigenstress
respectively\footnote{For reference, we can express with $\Mcal$ and
  $\Ncal$ both the surface vertical displacement due to an applied
  pressure
  $\overline\Mcal[p] = \Mcal[p\cdot \tens
  e_3]\big|_{\partial\body}\cdot\tens e_3$ and the residual vertical
  displacement
  $\bar u_3^p =
  \Ncal[\hooke:\tens\epsilon^p]\big|_{\partial\body}\cdot \tens
  e_3$.}. Their complete formulation and application in a discretized
setting is extensively discussed
in~\citep{frerot_fourieraccelerated_2019}. The coupling between the
elastic contact problem and the plasticity problem is done with a
relaxed fixed point
strategy~\citep{jacq_development_2002,frerot_fourieraccelerated_2019}.
The full implementation of the described solution method is also
freely available in \emph{Tamaas}.

\subsection*{Comparison: rough surface}\label{sec:comparison}
While both plasticity models are phenomenological, associated
plasticity is soundly grounded in experimental
observations~\citep{bui_etude_1969} as well as thermodynamic
principles~\citep{reddy_internal_1994,simo_computational_1998}, and
expresses a macroscopic picture of dislocation systems at the
micro-scale. This is not the case for the saturation models: they
depend on the observation that the mean contact pressure saturates for
spherical indentation~\citep{tabor_hardness_1951}, which has been
challenged by recent finite-element
simulations~\citep{song_elastic_2013,krithivasan_analysis_2007}.

We aim here to provide a direct comparison for a rough surface between
a perfectly plastic $J_2$ model and the saturation model. The rough
surfaces we use throughout this work are self-affine random surfaces.
Their power-spectrum density is defined as
\begin{equation}
  \phi(\tens q)\! =\! \begin{cases}
    C {\left(\dfrac{q_l}{|\tens q|}\right)}^{-2(H+1)} & q_l \leq |\tens q| \leq q_s,\\
    0 & \text{otherwise}
  \end{cases}
\end{equation}
where $q_l, q_s$ are the spatial frequencies associated to the long
cutoff wavelength $\lambda_l$ and short cutoff wavelength $\lambda_s$
respectively, while $H$ is the Hurst exponent. We define two surfaces
with Hurst exponent $H = 0.8$: $S_1$ has a rather narrow spectrum with
$L / \lambda_l = 3$ and $\lambda_l / \lambda_s = 3$; $S_2$ has a
broader spectrum and a larger system size, with $L / \lambda_l = 16$
and $\lambda_l / \lambda_s = 8$. The root-mean-square of slopes
$h'_\text{rms}$ is $0.07$ for $S_1$ and $0.1$ for $S_2$. Since $S_1$
is used for visual comparison, the discretization size $\Delta l$ is
chosen such that $\lambda_s / \Delta l = 3$, while a depth of $L / 5$
is represented with $64$ points (totaling $243\times 243\times 64$
points). $S_2$ is used to study crack nucleation at the interface, and
therefore has a finer discretization
$\lambda_s / \Delta l \approx 5.7$, with a depth $L / 5$ represented
with $32$ points (totaling $729\times 729 \times 32$). Naturally, the
saturated simulations do not account for volumetric behavior, so the
number of points are $243\times 243$ and $729 \times 729$ for $S_1$
and $S_2$ respectively.

The normal loads applied in both models are adimensionalized by
$W_0 = E^* L^2 h'_\text{rms}$. In elasticity, this normalization
collapses load ($W$) vs.\ true contact area ($A_c$) for different
values of $h'_\text{rms}$ (but the same spectrum
parameters)~\citep{bush_elastic_1975,hyun_finiteelement_2004}. We do
not intend here to modify the spectrum parameters but merely scale
$h'_\text{rms}$, which therefore becomes a non-dimensional measure of
surface summit amplitude. This is convenient to compare the results of
the $J_2$ and saturated models to an elastic reference, as we expect
the contact behavior to depend on the surface peak amplitude because
of plasticity.

\begin{figure*}[h]
  \centering \includegraphics{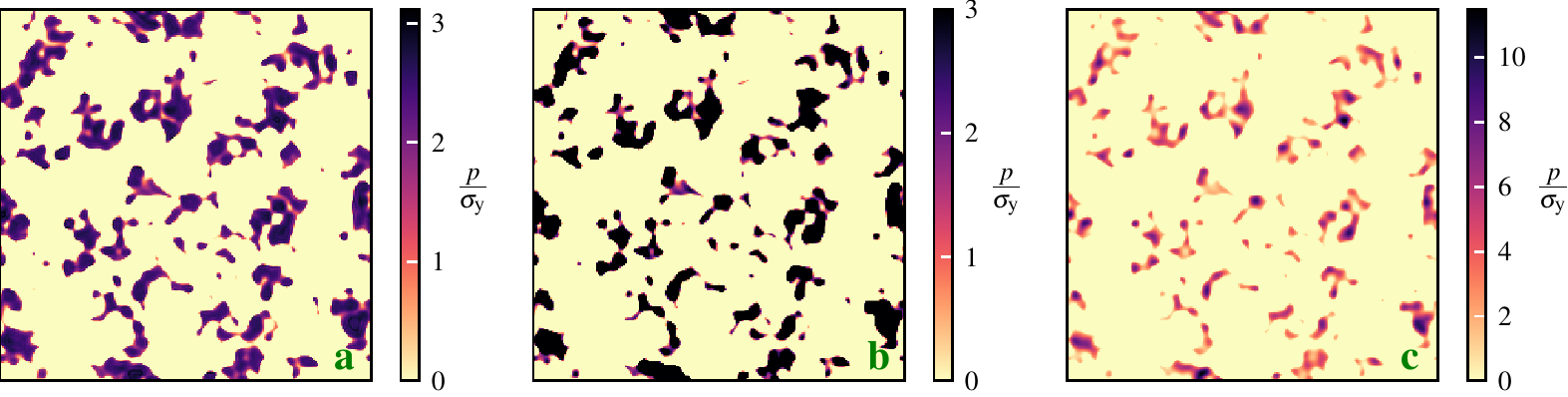}
  \caption{{\bfseries Pressure profiles for $J_2$ plasticity (a),
      saturation (b) and elasticity (c) models.} Although fig.\ (a)
    shows that the local pressure can exceed values of $3\sigmay$, the
    average pressure in contacts $\langle p \rangle = W / A_c$ is
    closer to $2\sigmay$, whereas the saturated model gives an average
    of $2.5\sigmay$ with large saturated portions of the
    micro-contacts. In this case, the normal load is $W / W_0$ is
    $6.5\cdot 10^{-2}$, and the saturated model predicts a contact
    area $20\%$ smaller than the $J_2$ prediction. As a result the
    connectivity of micro-contacts is different between the two
    models.}\label{fig:plastic_comparison}
\end{figure*}
\Cref{fig:plastic_comparison} shows the contact pressures for $J_2$
(\cref{fig:plastic_comparison}a) and saturated
(\cref{fig:plastic_comparison}b) plasticity, as well as elasticity
(\cref{fig:plastic_comparison}c), at the load
$W / W_0 = 6.5\cdot 10^{-2}$. The yield stress is
$\sigmay = 10^{-2} \cdot E$ and the surface used is $S_1$. The total
contact ratio is $25\%$ for $J_2$ plasticity, $20\%$ for saturation
and $15\%$ for elasticity, resulting in about $20\%$ error in the
contact area of the saturated model. Moreover, while the maximum
pressure in the $J_2$ model exceeds $3\sigmay$ (cf.
\cref{fig:plastic_comparison}a), the average pressure on
micro-contacts is closer to $2\sigmay$, which the saturation model
fails to capture with an average of $2.5\sigma_y$. Local features of
the contact patches also differ due to the three models being in
different contact stages.

\begin{figure}[t]
  \centering \includegraphics{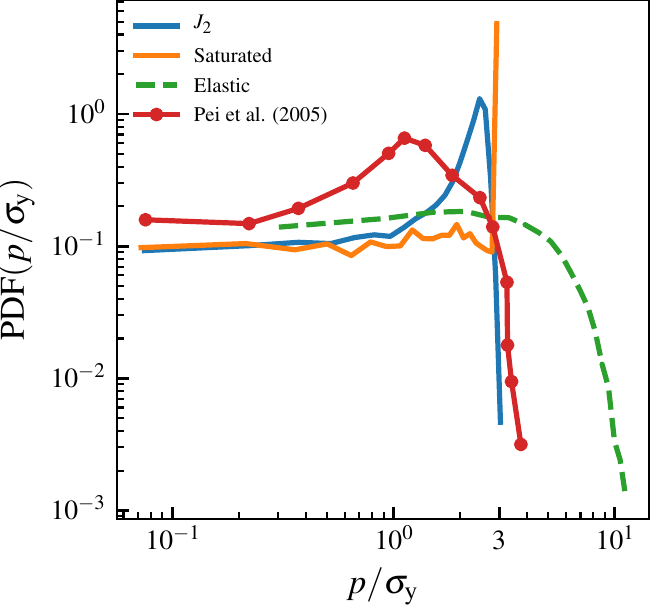}
  \caption{\textbf{Probability density function of surface pressures.}
    Neither the elastic nor the saturated models qualitatively
    reproduce the pressure distribution of the elastic-plastic model.
    As expected, the pressure distribution of the saturated model
    tends to a Dirac distribution for $p = p_m$, whereas the
    distribution for the elastic-plastic model tends to zero. Note
    that the results of \citet{pei_finite_2005} have been renormalized
    (cf.\ companion
    notebook~\citep{frerot_supplementary_2019a}).}\label{fig:pdf_saturated}
\end{figure}
\begin{figure}[t]
  \centering \includegraphics{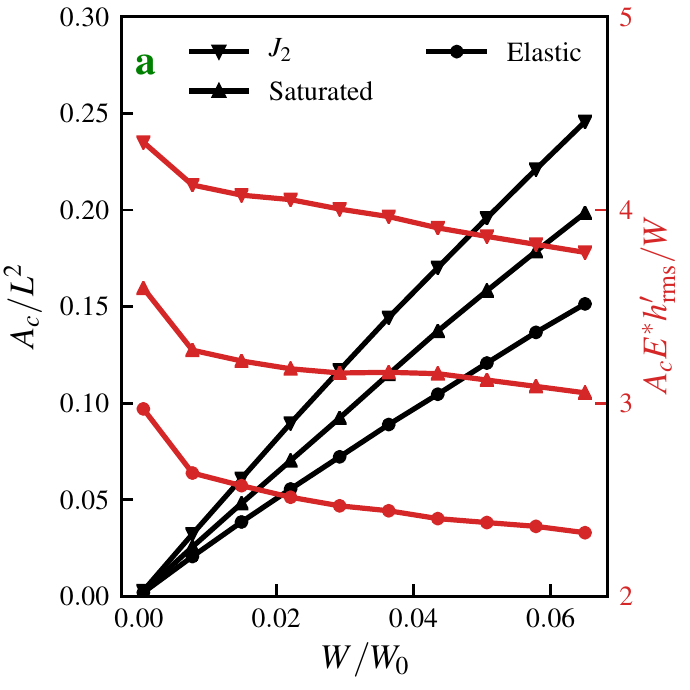}
  \includegraphics{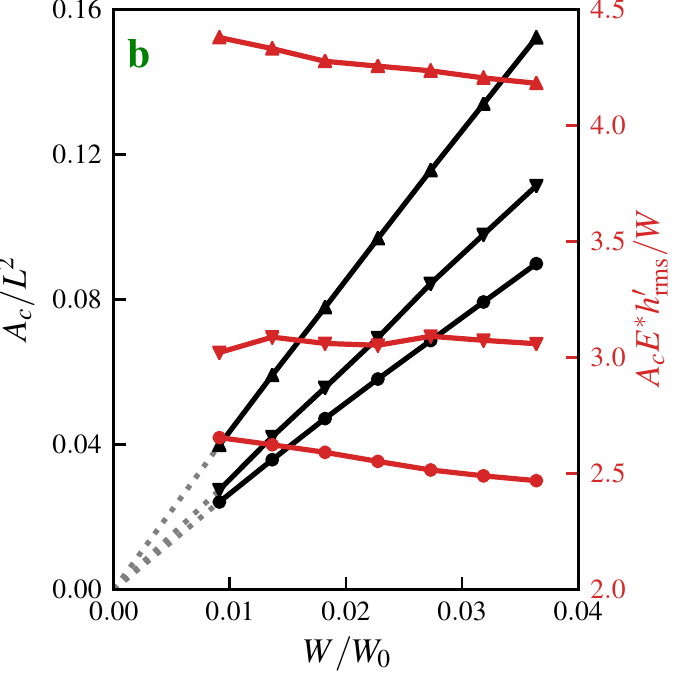}
  \caption{\textbf{Contact ratio and secant slope of the load/contact
      area relationship.} Figure (a) shows results for surface $S_1$
    and elastic perfectly-plastic materials. Figure (b) shows results
    for surface $S_2$ (which has a broader spectrum), and $5\%$
    hardening for the $J_2$ model. In fig. (a), the saturated model
    overestimates the true contact area, whereas it is underestimated
    in fig. (b). This suggests that adjusting the saturated model to
    quantitatively reproduce the true contact area may only be done on
    a case-by-case basis.}\label{fig:secant_slope}
\end{figure}

\Cref{fig:pdf_saturated} shows the probability density function of the
surface pressures for the three models, with the addition of reference
data from \citet{pei_finite_2005} (fig. 10a) which results from a
$J_2$ plasticity criterion used in a finite-element approach. The
features of the curve corresponding to the $J_2$ plasticity models are
not qualitatively reproduced by the saturation model: the peak at
$p = 2.5\sigmay$ is non-existent and as expected the distribution of
the saturated pressures tends to a Dirac at $p = p_m$ whereas the
$J_2$ distribution tends to zero. The difference between our results
and those of \citet{pei_finite_2005} can be explained by the
coarseness of the mesh they used, as well as different spectrum
parameters (e.g.\ $H = 0.5$ in their case).

\Cref{fig:secant_slope} shows the contact area ratio and the secant
slope of the load/contact area relationship, which is a quantity that
has been extensively studied in elastic contact
\citep{bush_elastic_1975,persson_theory_2001,yastrebov_infinitesimal_2015,campana_contact_2007,pastewka_contact_2014,yastrebov_role_2017,hyun_finiteelement_2004}.
\Cref{fig:secant_slope}a shows the result for surface $S_1$: as in
\cref{fig:plastic_comparison}, the saturated model underestimates the
true contact area compared to the $J_2$ plasticity model. All three
model nonetheless exhibit a similar behavior, which is not the case in
\cref{fig:secant_slope}b, where we show the same data for surface
$S_2$. Due to computational issues, the $J_2$ approach in this case
includes hardening with $E\hsub / E = 0.05$. This locally increases
the yield stress, but in our simulations the cumulated plastic strain
averaged over plastic zones never exceeds $1\%$, which means that the
yield stress only increases by $5\%$ on average. One can note that
this time the saturated model overestimates the true contact area,
meaning that the average pressure on contacts in the $J_2$ case is
larger than $3\sigmay$. This is consistent with the findings of
\citet{pei_finite_2005} and \citet{krithivasan_analysis_2007} with
elastic perfectly-plastic materials\footnote{We have $\langle p \rangle \approx 3.6\sigmay$, which is $20\%$ larger than the common $3\sigmay$. Since hardening only increases $\sigmay$ by $5\%$ on average, this difference in $\langle p \rangle$ cannot be explained by hardening alone.}. Unlike for the saturated and
elastic models, the secant slope of the $J_2$ model seems to stay
constant, as observed by \citet{pei_finite_2005}. This observation,
and in general the dependency of the contact area to load relationship
on surface and plasticity parameters warrants further investigations,
similar to \citet{pei_finite_2005}, with refined simulations, which
unfortunately go out of the scope of this work.

To study macroscopic quantities (such as the true contact area), the
saturation model is only suitable if one has reference data to fit
$p_m$. However, it fails to qualitatively reproduce local surface
quantities (as can be seen in
\cref{fig:plastic_comparison,fig:pdf_saturated}), as well as global
quantities in certain cases (e.g.\ \cref{fig:secant_slopes}b), in
addition to providing no information on the complete stress state at
and below the contact surface. The saturation model may be useful in
applications where quantitative errors in the contact area magnitude
or its topography may be accepted, only to obtain approximate qualitative
relations. It can however give unequivocally wrong results when local
stress based quantities drive the phenomenon one wishes to study. We
shall see in the next section that this can be the case in wear
modeling when we consider the crack nucleation process in an
elastic-plastic rough contact.


\section{Crack nucleation in rough surface
  contact}\label{sec:crack_rough}
Most of the atomistic investigations of adhesive wear processes use
geometries that contain stress
concentrators~\citep{aghababaei_critical_2016,milanese_emergence_2019,brink_adhesive_2019},
such that in their model system the debris formation is only
controlled by the Griffith energy balance. However, without a
defect/stress concentration, a crack described by linear elastic
fracture mechanics cannot nucleate. In a half-space geometry, with
plasticity constitutive behavior, no such concentration can exist. For
that reason, it is necessary to introduce a critical nucleation
tensile stress $\sigma_c$.

In linear elasticity the stresses are unbounded and depend linearly on
the applied load. The picture however changes in plasticity, as one
can expect the tensile stress to saturate, possibly below $\sigma_c$,
preventing crack nucleation altogether. \Citet{brink_adhesive_2019}
have also shown that the resistance to shear of the contact junction
plays a fundamental role in the wear particle formation. From a stress
perspective, there is a competition between $\sigma_c$ and $\tau_j$
(the junction shear resistance) for the formation of a crack: if the
junction is strong enough, the maximum tensile stress may, under
conditions depending on $\sigmay$, reach $\sigma_c$ and nucleate a
crack. Conversely, if the junction is weaker, its slip may prevent the
tensile stress from reaching $\sigma_c$.

We investigate the interplay of these effects in the contact of the
rigid self-affine rough surface $S_2$ (defined previously) with a
solid that we have discretized with $729\times729$ points for the
elastic and saturation models (with $\sigmay / E = 10^{-2}$), and
$729\times 729\times 32$ for the $J_2$ model ($\sigmay / E = 10^{-2}$,
$E\hsub / E = 5\cdot 10^{-2}$ as in the previous section). The applied
normal mean pressure varies between $10^{-2} h'_\text{rms}E$ and
$8\cdot 10^{-2} h'_\text{rms} E$ for the elastic case and between
$10^{-2} h'_\text{rms}E$ and $4\cdot 10^{-2} h'_\text{rms} E$ for the
saturation and $J_2$ models. Since we investigate the effect of the
junction shear strength, we also apply a shear stress on the contacts
at constant normal load. There is a simple correspondence between the
applied shear stress and the junction strength $\tau_j$: any shear
loading larger than $\tau_j$ should not modify the stress state of the
system since all contacts should be slipping and the stresses should
not increase. We therefore interchangeably refer to the applied shear
and the junction strength as $\tau_j$. In linear elasticity, the
application of a constant shear stress on a patch of the surface
creates a stress singularity at the edge of the patch because the
derivative of the surface tangential displacement
diverges~\citep{menga_surface_2019}. In a physical system, a small
amount of slip and rearrangement of the solids would occur at the edge
of the contact junction so that the shear stress carried should be
reduced on this zone. We therefore regularize\footnote{Regularization
  is done by convolution with a function of the form
  $\phi_\epsilon(r) = \exp(-1 / (1 - (r/\epsilon))) /
  \epsilon^2$~\citep{david_equations_2015}.} the constant shear
distribution over a transition zone of width $\epsilon_\tau$ small
compared to the smallest surface wavelength
($\epsilon_\tau \approx \lambda_s/3$), removing any numerical
discrepancy due to the singularity.

We call a crack nucleation site a connected zone of the surface where the
largest eigenvalue of the Cauchy stress tensor $\sigma\I$ is larger than
$\sigma_c$. Although the wear particle formation process is deterministic, the
inherent randomness of the rough surfaces makes the process epistemically
random~\citep{frerot_mechanistic_2018}. We apply a similar concept here and
study the probability that a given contact nucleates a crack. Although we focus
here on crack nucleation, and do not model crack propagation, we assume that the
propagation of a crack has a minimal influence on the nucleation of other
cracks. This assumption holds for small true contact areas, where the
micro-contacts are dilute, which would correspond to a mild wear
regime\footnote{It is possible that this assumption holds for large contact area
  ratios~\citep{sevostianov_effective_2012}, but current numerical methods do
  not allow for detailed contact simulations with explicit crack propagation
  modeling to verify if this is the case.}. We shall see that despite this
assumption, our model still includes elastic interactions between contacts and
plastic zones, which have an effect on the crack nucleation sites in close
proximity, particularly when shear is applied.

\begin{figure}[t]
  \centering \includegraphics{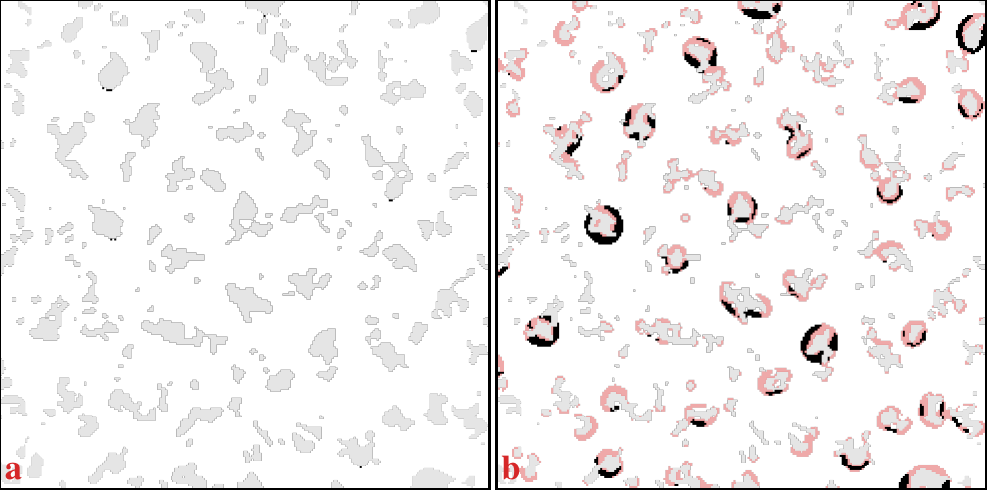}
  \caption{\textbf{Crack nucleation sites for the saturation pressure
      (a) and $J_2$ plasticity (b) models.} An ``upwards'' shear is
    applied on each contact. The true contact area is shown in light
    grey, the plastic zones in red and the crack nucleation sites in
    black. We can see that the $J_2$ model has more crack nucleation
    sites than the saturation model. Since both models give
    approximately the same true contact area, this discrepancy must be
    due to plastic residual deformations which are not represented in
    the saturation approach. Figure (b) shows that there can be
    multiple crack nucleation sites per
    contact.}\label{fig:crack_sites}
\end{figure}

Because of the complex topography of the micro-contacts, there is no one-to-one
correspondence between crack nucleation sites and contacts and the probability
that a given contact nucleates a crack cannot be explicitly evaluated. This can
be seen in \cref{fig:crack_sites}, where we show in grey the true contact area,
in red the plastic zones and in black the crack nucleation sites (i.e.\ the
zones where $\sigma\I > \sigma_c$) for $\sigma_c = 0.1 h_\text{rms}'E^*$ and
upwards applied shear stress $\tau_j = 2\cdot 10^{-3} E$.
\Cref{fig:crack_sites}a shows the result for the saturation plasticity model and
\cref{fig:crack_sites}b shows the $J_2$ plasticity model. One can easily see on
the latter that a single micro-contact may have several crack sites. We also
recognize the expected crescent shape crack nucleation on some contacts.
Finally, the number of cracks is larger in the $J_2$ model than in the
saturation plasticity model. Since the contact areas predicted by both models
are essentially the same, this difference can only be explained by the residual
plastic deformations, which are not modeled in the saturation plasticity
approach.

Because we still want to investigate a quantity akin to the
probability that a contact nucleates a crack, we propose an
adimensional measure called the crack nucleation likelihood (CNL)
given by $A_0\cdot n_\text{crack} / A_c$, where $A_0 = L^2$ is the
apparent contact area and $n_\text{crack}$ is the number of crack
nucleation sites (i.e.\ the number of connected black zones in
\cref{fig:crack_sites}). The CNL is conceptually a normalization of
the number of crack nucleation sites by the density of contacts and
relates to the probability of crack nucleation at a contact (this
relationship will be detailed later on in this article).

\begin{figure*}
  \centering \includegraphics{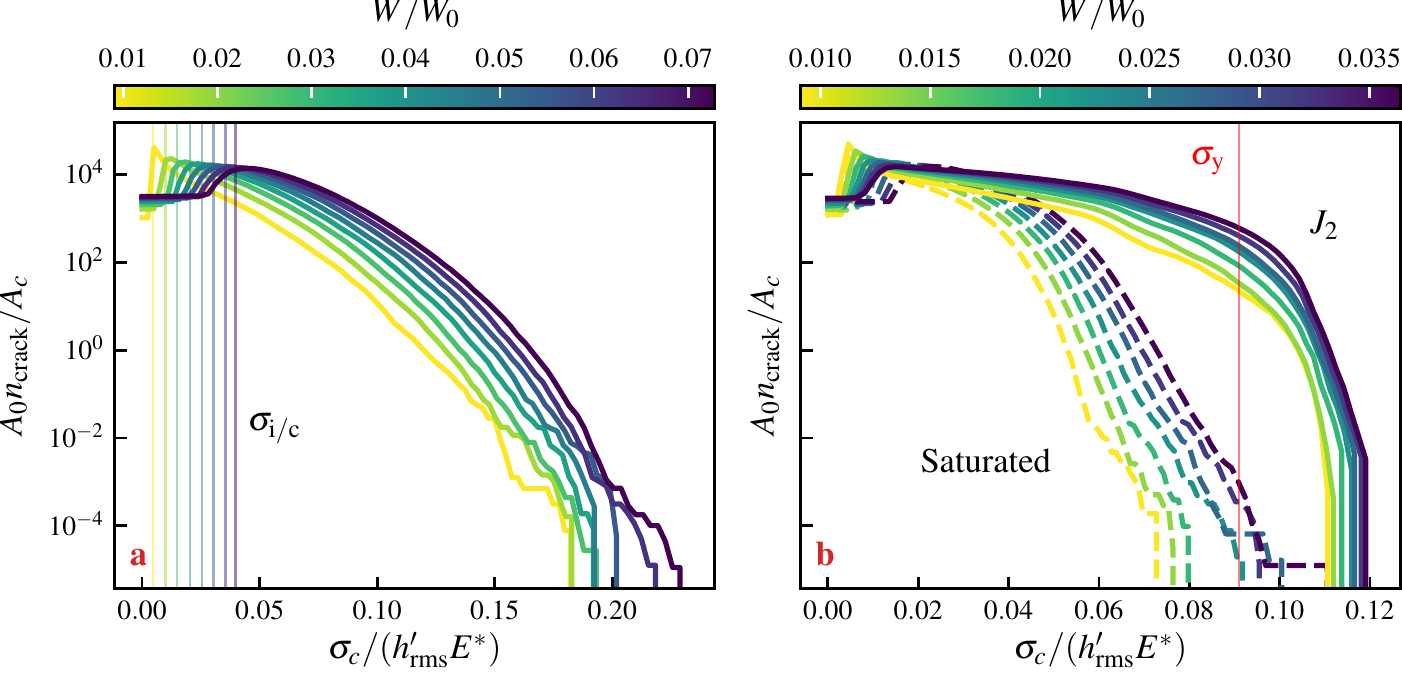}
  \caption{\textbf{Crack nucleation likelihood (CNL) as a function of
      $\sigma_c$ and normal load.} Figure (a) shows the results for an
    elastic contact, while (b) shows the saturation and $J_2$
    plasticity results. In (a) the CNL curves are uniformly shifted to
    the right when the load is increased, indicating an exponential
    increase in the CNL.\@ The normalization of $n_\text{crack}$ by
    the true contact area makes explicit that this increase is due to
    stronger elastic interactions between contacts. The magnitude of
    the shift is given by the most frequent value of $\sigma\I$ called
    $\sigma\ic$. The two plasticity models in (b) have drastically
    different behavior: the crack nucleation is much more likely in
    the $J_2$ approach because of plastic residual stresses, and the
    CNL experiences a faster decay for values of
    $\sigma_c > \sigmay$.}\label{fig:density_unnormalized}
\end{figure*}

\begin{figure*}
  \centering \includegraphics{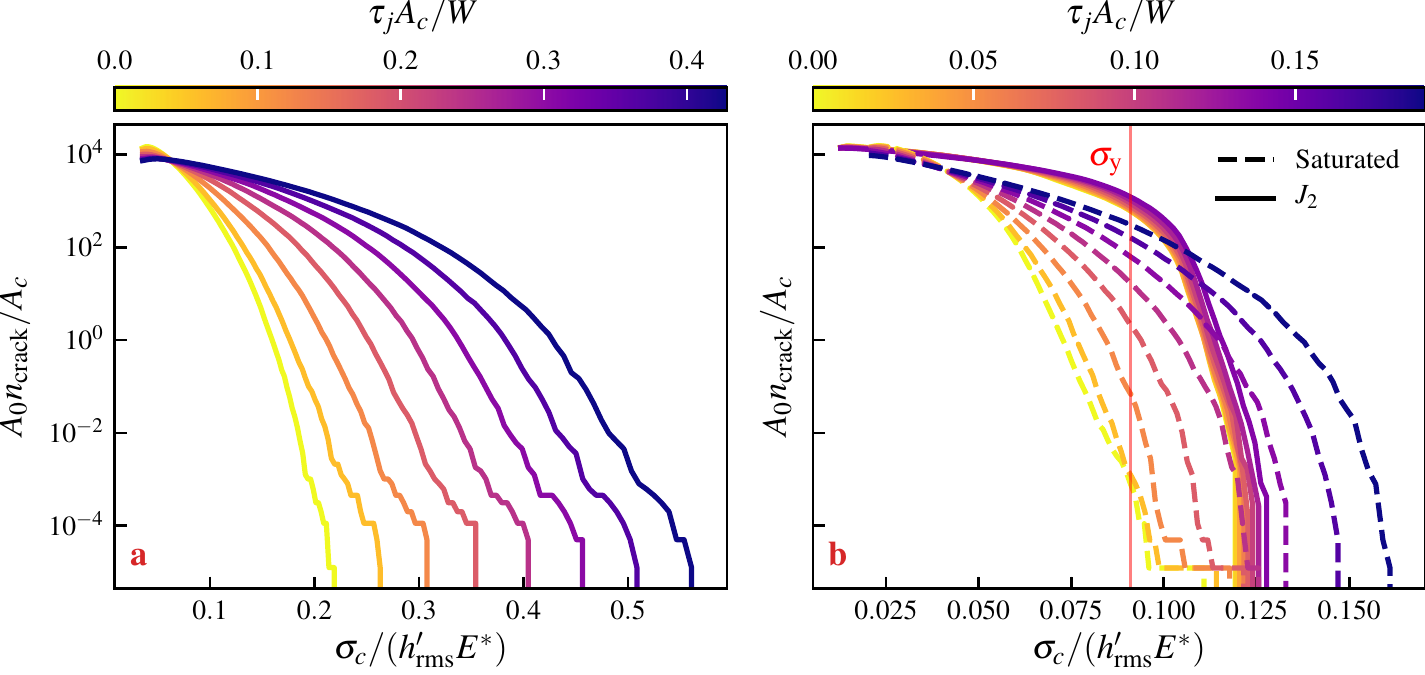}
  \caption{\textbf{Crack nucleation likelihood (CNL) as a function of
      $\sigma_c$ and junction strength.} As for
    \cref{fig:density_unnormalized}, (a) shows the elastic model and
    (b) the plastic models. Unlike \cref{fig:density_unnormalized}a,
    the CNL for the elastic model is \emph{scaled} to the right when
    the junction strength increases. The same can be said of the
    curves corresponding to the saturation plasticity in (b), but not
    of the $J_2$ curves, which are relatively insensitive to changes
    in $\tau_j$. This is due to plasticity preventing increases in
    $\sigma\I$.}\label{fig:density_shear_unnormalized}
\end{figure*}

\Cref{fig:density_unnormalized,fig:density_shear_unnormalized} show
the CNL as a function of $\sigma_c$ when the normal load and the
applied shear stress/junction strength are respectively varied, for an
elastic (a), a saturated and a $J_2$ contact (b). On
\cref{fig:density_unnormalized}a, the CNL curves are shifted to the
right with larger normal loads. This means that for a fixed value of
$\sigma_c$, the CNL increases exponentially when the load is
increased. The vertical lines indicate a quantity $\sigma\ic$, which
is the stress for which the probability density function of $\sigma\I$
on the whole surface is maximum: in other words it is the most
frequent stress value, and is typically found between contacts (hence
the term ``inter-contact stress''). The horizontal shift in the CNL
curves corresponds to $\sigma\ic$. Since the latter depends on the
spatial proximity of contacts, the CNL must depend on elastic
interactions between contacts. \Cref{fig:density_unnormalized}b shows
that the two plasticity models have widely different behavior: the
crack nucleation likelihood is much higher in the case of $J_2$
plasticity, but also decays faster for values of $\sigma_c > \sigmay$.
It is surprising that despite being plastic with some hardening, the
$J_2$ model is more likely to lead to surface cracks than the
saturation model. This may be a first step towards resolving the
paradox highlighted in introduction.

\Cref{fig:density_shear_unnormalized} shows the CNL when the junction
strength increases (for the last normal load of
\cref{fig:density_unnormalized}). Unlike previously, the elastic
results are not shifted to the right when $\tau_j$ increases but are
instead \emph{scaled} rightward. The same happens with the saturation
plasticity model, whereas the $J_2$ CNL seems relatively unaffected by
$\tau_j$. This is due to plastic deformations having reached the
contact surface (cf.\ \cref{fig:crack_sites}) and preventing an
increase of $\sigma\I$ as fast as the elastic and saturation models,
although hardening still allows some increase at a lower rate. In
order to rationalize these results and provide evidence for the
conclusions we have induced from our multi-asperity simulations, we
now study the crack nucleation process for a single asperity.

\section{Single asperity crack nucleation}\label{sec:single}
To investigate the effect of plasticity in the competition between
$\sigma_c$ and $\tau_j$, we simulate a spherical indenter of radius
$R$ pushed onto an elastic-perfectly-plastic solid. The resulting
contact junction is then subjected to a shear distribution (with
$\epsilon_\tau = R/64$), and the principal tensile stress $\sigma\I$
at the surface is recorded.

\begin{figure}[t]
  \centering \includegraphics{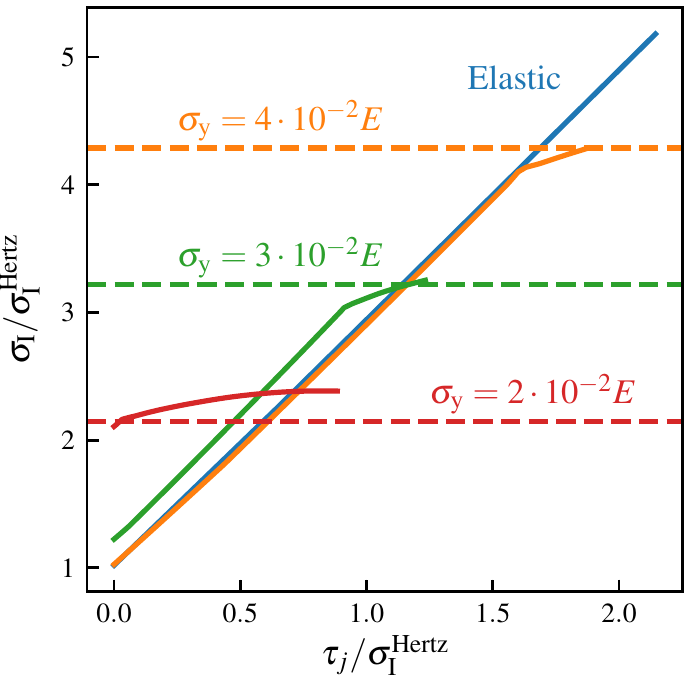}
  \caption{\textbf{Maximum tensile stress as a function of
      applied shear stress across single asperity contact with $J_2$
      plasticity.} The loading curves consist of two parts: an initial
    linear loading and a non-linear saturation of $\sigma\I$. Although
    the initial slope is linear, the loading is not elastic as plastic
    deformations still evolve in the system. Their influence on the
    maximum tensile stress at the surface is however minimal. A
    sufficient shear stress may eventually create plastic strain at
    the surface, which causes a transition in the loading curve.
    Moreover, given a low enough yield stress, the initial indentation
    may cause surface yield, as seen for the curve where
    $\sigmay/E = 2\cdot 10^{-2}$.}\label{fig:tensile_single}
\end{figure}
\Cref{fig:tensile_single} shows the maximum tensile stress $\sigma\I$
as a function of the applied shear stress $\tau_j$ across the contact.
The different curves correspond to different yield stresses, with the
dashed lines indicating the value of $\sigmay$ for reference. Stresses
here are normalized by the maximum tensile stress in Hertz contact
$\sigma\I^\text{Hertz} = (1 - 2\nu) p_0 / 3$, with $p_0$ being the
maximum hertzian contact pressure~\citep{johnson_contact_1985a}. We
can observe that the initial tensile stress (without applied shear)
depends on the amount of plastic deformation: if $\sigmay$ is
decreased (or conversely the applied load increases), the initial
tensile stress at the edge of the contact is higher. This is due to
the residual stresses created by the plastic deformations that
accommodate the indentation: the localized nature of the plastic
strains causes the unloaded equilibrium position to not be
stress-free. The additional stresses are tensile and add to the stress
on the contact rim. In the case of $\sigmay / E = 2\cdot 10^{-2}$, the
plastic zone has reached the surface and the von Mises stress at the
edge of contact has reached $\sigmay$ by indentation alone (not shown
here).

If the von Mises stress at the surface is below $\sigmay$, the
application of a shear stress will cause an elastic loading phase,
offset by the initial $\sigma\I$ value, as seen for the higher values
of $\sigmay / E$. The loading continues until $\sigma\I$ reaches
values close to $\sigmay$, as indicated by the dashed lines. Since the
stress state at the edge of contact is triaxial, the maximum value
$\sigma\I$ can reach is not $\sigmay$, as is seen for the most plastic
case. After a certain point, \cref{eq:implicit_plasticity} becomes
unsolvable because a plastic failure mechanism
develops~\citep{drucker_soil_1952}: we supposed that further loading
will not increase the value of $\sigma\I$. Of course, for a hardening
material $\sigma\I$ should not saturate and instead increase further
at a lower rate.

\begin{figure*}[t]
  \centering \includegraphics{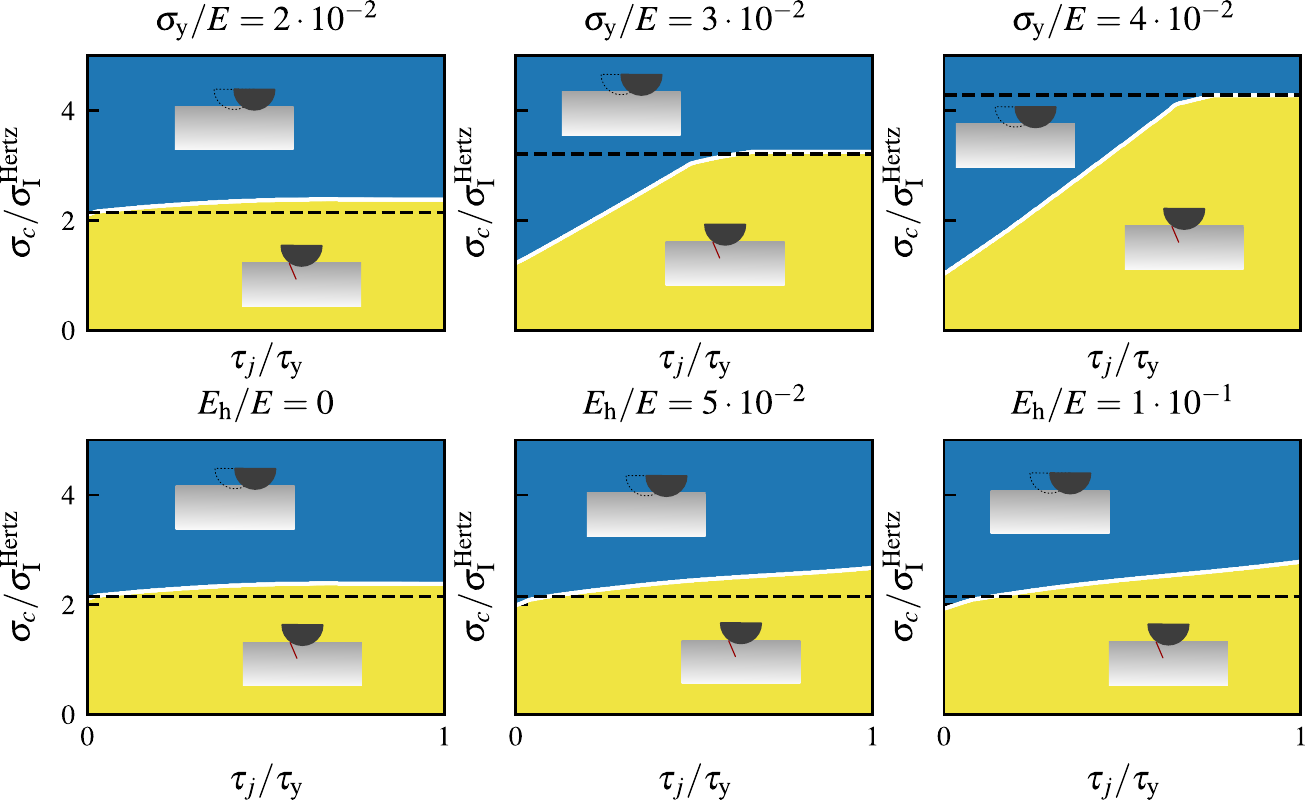}
  \caption{\textbf{Failure regimes for a sheared spherical
      indentation.} For the first row of graphs, the ratio
    $\sigmay / E$ is varied for a perfectly plastic material. For the
    second row, the hardening ratio $E\hsub / E$ is varied for a yield
    ratio of $\sigmay / E = 2\cdot 10^{-2}$. On each graph the dashed
    line shows the yield stress. The white curve marks the transition
    between failure driven by slip rupture of the junction and failure
    by crack nucleation. The competition between the junction strength
    $\tau_j$ and the critical stress $\sigma_c$ is influenced by
    $\sigmay$ because of the saturation effect shown in
    \cref{fig:tensile_single}. While plasticity gives a failure
    mechanism independent of $\tau_j$, hardening allows the tensile
    stress to grow past the initial yield limit, giving a linear
    transition between failure mechanisms.}\label{fig:crack_phases}
\end{figure*}
\Cref{fig:crack_phases} shows the competition between the junction
strength $\tau_j$ and the tensile strength $\sigma_c$ for different
values of $\sigmay$ (first row no hardening, i.e.\ $E\hsub / E = 0$)
and $E\hsub$ (second row with $\sigmay / E = 2\cdot 10^{-2}$) the
hardening modulus. The zones below the white curve show for which
values of $(\sigma_c,\tau_j)$ a crack may nucleate, and the zones
above show when the interface breaks (i.e.\ slips) before crack
nucleation. The dashed lines indicate the values of $\sigmay$. We can
see that for materials with high yield there is an affine boundary
between the two mechanisms, which shows their competition. Of course
materials with $\sigma_c < \sigmay$ are not plastic, so such
transition can only happen for brittle materials. It does however
exist for hardening materials, as opposed to perfectly plastic ones.
With no hardening, the failure mechanism is purely determined by the
value of $\sigma_c / \sigma\I^\text{Hertz}$, which depends on the
applied load.

With this single-asperity analysis, we have explained why the crack
nucleation likelihood is higher in the $J_2$ model for no applied
shear stress: the plastic residual deformations cause tensile stresses
which combine with the contact stresses and increase $\sigma\I$, thus
increasing the CNL.\@ This does not occur in the saturated model
because it ignores plastic residual deformations in the stress
computation. We have also explained why the $J_2$ CNL is relatively
insensitive to the applied shear/junction strength: plasticity has a
saturation effect on $\sigma\I$: when the system is sheared, the rim
of contacts is in the plastic regime, and the increase in $\sigma\I$
is purely driven by hardening, which in the case of
\cref{fig:density_shear_unnormalized} is only 5\% of the Young's
modulus. In order to rationalize the other aspects of the CNL
highlighted by
\cref{fig:density_unnormalized,fig:density_shear_unnormalized}, we
resort to a statistical model for multi-asperity contact.

\section{Multi-asperities}\label{sec:multi}
One can now apply a statistical approach to estimate the proportion of
contacts that nucleate cracks in a multi-asperity setting. We
thereafter use a Greenwood--Williamson (GW)
model~\citep{greenwood_contact_1966} with an exponential distribution
of asperity heights to obtain simple, qualitative, analytical results
to help rationalize the data of
\cref{fig:density_unnormalized,fig:density_shear_unnormalized}. Since
the asperities are randomly distributed (all with the same radius
$R$), $\sigma\I^\text{Hertz}$ becomes in turn a random variable. There
is however a significant difference between the single asperity case
we have studied and the multi-asperity setting. Because of elastic
interactions, the tensile stress at the edge of a contact depends on
the proximity and magnitude of the neighboring contacts. In a
traditional GW approach, contacts are independent of each other. We
assume this is the case, but that the stress state is determined by
the local contact with an additional contribution
$\sigma_\mathrm{i/c}$, the inter-contact stress, determined from the
neighboring contacts. The radial stress outside the area of a single
contact of radius $a$ is given by~\citep{johnson_contact_1985a}:
\begin{equation}\label{eqn:sigma_r}
  \sigma_r(r) = \frac{1 - 2\nu}{3}\cdot\frac{a^2}{r^2}p_0
  = \kappa\frac{sR}{r^2}{\left(z^* - h\right)}^\frac{3}{2},
\end{equation}
with
\begin{equation}\label{eqn:kappa}
  \kappa := \frac{2(1-2\nu)E^*}{3\pi}\cdot\sqrt{\frac{s}{R}},
\end{equation}
where $r$ is the euclidean distance from the contact center,
$z^* = z / s$ is the asperity height random variable normalized by the
standard deviation of heights $s$ and $h$ is the normalized surface
approach. We assume a spatial asperity density $\eta$ and a contact
density $\eta_c = \eta e^{-h} = A_c / (A_0 \pi sR)$
\citep{greenwood_contact_1966}, with $A_c$ being the true contact
area. To compute the inter-contact stress $\sigma\ic$, we assume the
stress state outside each contact is given by the mean of
\cref{eqn:sigma_r}, averaged over contacting asperities, and we
compute the largest stress eigenvalue at the center of a series of
circles whose diameters are multiples of $d_c = 1/\sqrt{\eta_c}$ which
is the characteristic distance between contacts. The calculation
process is detailed in \cref{app:sigma_ic} and leads to the following
expression:
\begin{align}\label{eqn:sigma_ic}
  \sigma\ic = 3\kappa\frac{\xi}{\sqrt{\pi}}\cdot\frac{A_c}{A_0},
\end{align}
with $\xi \leq \sqrt{3\zeta(3)} \approx 1.9$ and $\zeta$ is the
Riemann Zeta function. We use this upper bound for $\sigma\ic$ in the
rest of this work. We can see that $\sigma\ic$ depends linearly on the
contact ratio, and therefore is linear with the load. We suppose that
$\sigma\ic$ acts as a ``background'' stress, and that the maximum
tensile stress is the sum of the local contact tensile stress
$\sigma\I^\mathrm{Hertz}$ and the inter-contact stress. We can
therefore quantify the probability that a contact nucleates a crack:
\begin{align}
  P_\mathrm{crack} & = P\left(\frac{\sigma_c}{\sigma\I^\mathrm{Hertz} + \sigma\ic} < \omega(\tau_j) \ \Big|\ z^* - h \geq 0\right)\nonumber\\
  \label{eqn:p_crack}& = \exp\left(-{\left(\frac{\sigma_c/\omega(\tau_j) - \sigma\ic}{\kappa}\right)}^2\right),
\end{align}
where $\omega$ is the function describing the failure mechanism
transition (white line in \cref{fig:crack_phases}). The calculation
details of \cref{eqn:p_crack} can be found in \cref{app:p_crack}.

\begin{remark}\label{rem:cnl}
  The crack nucleation likelihood $A_0\cdot n_\text{crack} / A_c$ and
  $P_\text{crack}$ are related in our GW approach: indeed
  $n_\text{crack} = P_\text{crack}\eta_c A_0 = P_\text{crack} A_c /
  \pi sR$, cf.\ \citep[p. 303]{greenwood_contact_1966}.
\end{remark}
\begin{remark}\label{rem:scaling}
  The quantity $\sigma_c / \omega(\tau_j) - \sigma\ic$ strikingly
  explains the features of
  \cref{fig:density_unnormalized,fig:density_shear_unnormalized} for
  the elastic model. In \cref{fig:density_unnormalized}, the curves
  are shifted to the right as the load increases, which is apparent in
  \cref{eqn:sigma_ic,eqn:p_crack}: $\sigma\ic$ increases linearly with
  the load, and thus causes a rightward shift in the graph of the
  CNL.\@ Similarly, as $\omega$ is linear in $\tau_j$,
  $P_\text{crack}$ is scaled horizontally, which can also be seen in
  the CNL on \cref{fig:density_shear_unnormalized}.
\end{remark}
\begin{remark}
  When $\sigma_c / \omega(\tau_j) = \sigma\ic$ the probability is one,
  meaning that all contacts, regardless of size, nucleate cracks; in
  other words the whole surface should be cracking in a catastrophic
  breakdown. This does not happen in practice, as the normal loading
  process should nucleate and propagate cracks at single asperities
  before the breakdown is reached, thus relaxing the tensile stresses
  in the system.
\end{remark}

\subsection*{Comparison to a rough surface}
We wish to assess the validity of the above developments with
simulations of self-affine rough surface contact. Because of the
simplifying assumptions of a GW model, we do not hope to establish a
quantitative agreement, especially since the asperity curvature is not
unequivocally defined on a self-affine rough
surface~\citep{nayak_random_1971}. Instead, we will focus on the
qualitative relations between $P_\mathrm{crack}$, $\sigma_c$,
$\sigma\ic$ and $\tau_j$ highlighted above.

\begin{figure}
  \centering \includegraphics{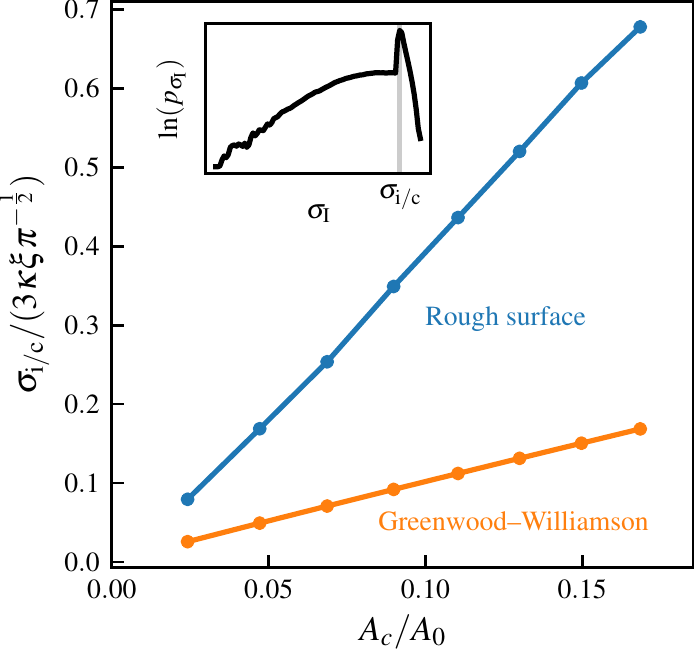}
  \caption{\textbf{Inter-contact stress as a function of contact
      ratio.} Inset shows the probability density function of the
    largest stress eigenvalue at the surface. The stress value
    corresponding to the peak in the probability density is defined as
    the inter-contact stress $\sigma\ic$. We see that in the rough
    surface simulation and the Greenwood--Williamson model $\sigma\ic$
    evolves linearly with the contact ratio. The value of $\kappa$ is
    estimated from the mean curvature of contacting summits in the
    rough surface.}\label{fig:intercontact}
\end{figure}

\paragraph{Elasticity Results} We first consider $\sigma\ic$ for a
rough surface. Recall that the inter-contact stress in a rough contact
is the most frequent value of $\sigma\I$, i.e.\ the value for which
the probability density function $p_{\sigma\I}$ of the surface tensile
stress is maximum. This is illustrated in the inset of
\cref{fig:intercontact}. In the latter, we plot the evolution of
$\sigma\ic$ for the elastic rough contact defined previously and for
\cref{eqn:sigma_ic}. We can observe that both curves behave linearly
with the contact area ratio, with different slopes. To compute the
value of $\kappa$, which depends on $\sqrt{s / R}$, cf.
\cref{eqn:kappa}, we have used Nayak's
approach~\citep{nayak_random_1971} to estimate the mean curvature
radius of the zones of the rough surfaces in contact, i.e.\
$R = \sqrt{3/m_4}/{I(z_c^*)}$, where $m_4$ is the fourth moment of the
surface spectrum, $z_c^*$ is the normalized height of the surface in
contact and $I$ is a function defined in~\citep{nayak_random_1971}. As
can be seen in \cref{fig:intercontact}, the slope of the GW curve is
approximately constant, showing the weak dependency on $z_c^*$. Note
that although the values of $R$ and $A_c /A_0$ for the GW model are
informed from the rough surface simulation, there is no fit parameter
to the GW prediction. The lack of quantitative agreement between the
theoretical approach and the rough contact simulation shows the
prediction limit of asperity-based models.

\begin{figure*}[t]
  \centering \includegraphics{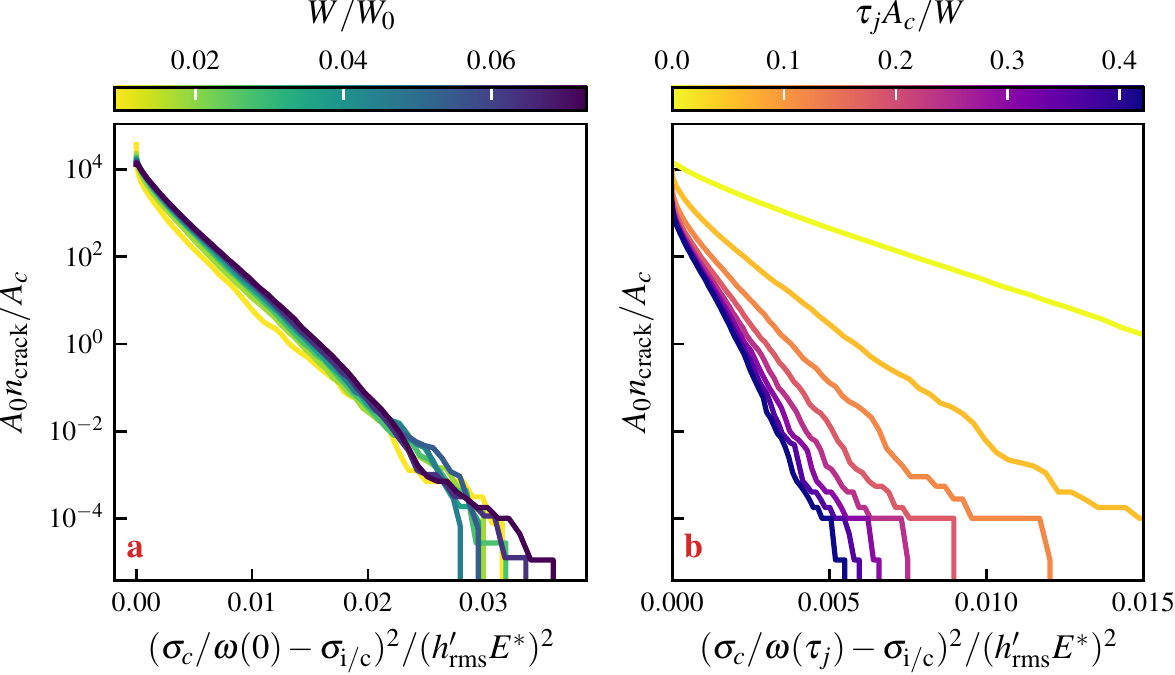}
  \caption{\textbf{Crack nucleation likelihood as a function of
      re-normalized critical stress, normal load and junction shear
      strength.} The material in contact behaves elastically. We
    normalize by $h'_\text{rms}E^*$ instead of $\kappa$ because of the
    difficulty of defining asperity curvature on a rough surface. One
    can see on figure (a) that the curves corresponding to different
    normal loads are collapsed on a straight line, showing that the
    CNL does indeed follow the scaling established in
    \cref{eqn:p_crack}. When the strength of the junction is taken
    into account (figure (b)), or, equivalently, if a shear stress is
    applied, we observe qualitative deviations from
    \cref{eqn:p_crack}. There is a \emph{decrease} in crack event
    density due to the interference of close contacts, which tends to
    unload the tensile stresses at the trailing edge of leading
    contacts.}\label{fig:density}
\end{figure*}

As previously mentioned, $P_\text{crack}$ is not directly measurable
on a rough contact interface (cf.\ \cref{fig:crack_sites}). However,
the crack nucleation likelihood acts as an alternative measure for
$P_\text{crack}$, cf.\ \cref{rem:cnl}. \Cref{fig:density}a, confirms
that this is indeed the case and we find the squared exponential
dependency predicted by \cref{eqn:p_crack}, with all curves collapsed
due to the shift caused by $\sigma\ic$ (recall that $\omega(0) = 1$).
\Cref{fig:density}b, on the other hand, shows that the CNL does not
follow \cref{eqn:p_crack} for non-zero $\tau_j$. While each curve
remains close to a straight line, they do not overlap, but seem to
converge to a master curve. More strikingly, the CNL \emph{decreases}
as $\tau_j$ increases, meaning that $\omega(\tau_j)$, which was
computed directly from the data of \cref{fig:tensile_single},
over-normalizes the data. This is again due to interactions between
asperities. For our ``single-asperity'' analysis, because of
periodicity, we in fact consider many interacting asperities on a
square lattice, each separated by a distance $L$. When shear is
applied, a positive $\sigma\I$ is created at the trailing edge of the
contact and a negative $\sigma\I$ appears at the leading edge. Because
the periodic images are equidistant and far apart, they weakly affect
the stress distribution in the vicinity of the contact. However, when
two contacts are close to each other, creating local anisotropy, the
inter-contact stress distribution of each asperity is compensated by
the other. The positive peak in $\sigma\I$ at the trailing edge of one
contact is then reduced, thereby reducing $P_\text{crack}$ as seen in
\cref{fig:density}b. This phenomena is akin to the crack shielding
mechanism uncovered by \citet{aghababaei_asperitylevel_2018}.

\begin{figure}
  \centering \includegraphics{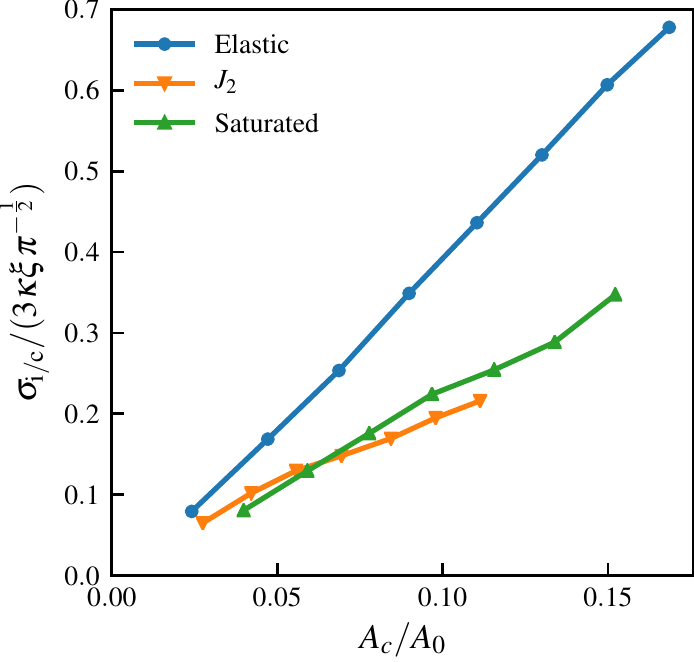}
  \caption{\textbf{Inter-contact stress as a function of contact ratio
      for the saturation and $J_2$ plasticity models.} While the
    curves do not match the analytical GW approach, their slopes are
    smaller than in the elastic case. The $J_2$ model shows the
    smallest slope, indicating that the stresses due to plastic
    residual deformations have an influence on the inter-contact
    stress and actually reduce it compare to the underlying elastic
    stresses of the saturation pressure model.
  }\label{fig:plastic_sigma_ic}
\end{figure}

\paragraph{Plasticity Results} As for the elastic model, we compare
the inter-contact stress computed from the plastic rough contact
simulations to our GW approach. \Cref{fig:plastic_sigma_ic} shows the
results for both the saturation and $J_2$ plasticity. Compared to the
elastic results, the slope of the plastic models is smaller. The $J_2$
plasticity model has the smallest slope, showing that residual plastic
deformations play a role in the inter-contact stress. One should note
that the plastic model in \cref{fig:plastic_sigma_ic} includes
hardening, hence the reduced contact ratio. It seems both models still
give a linear dependency of $\sigma\ic$ on the contact ratio, although
some more data may be required to draw an affirmative conclusion in
this regard.

\begin{figure}
  \centering \includegraphics{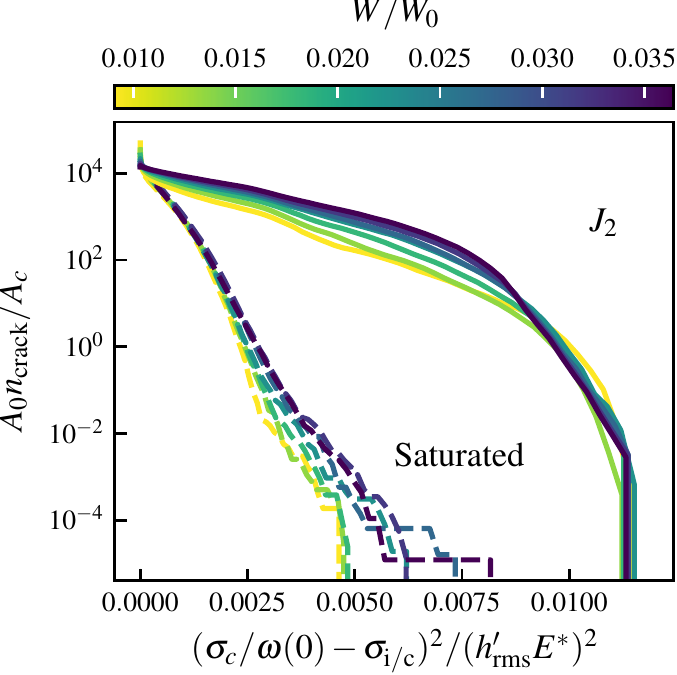}
  \caption{\textbf{Crack nucleation likelihood as a function of
      re-normalized critical stress and normal load for the saturation
      and $J_2$ plasticity models.} The pressure saturation model
    reproduces a scaling similar to the elastic case in
    \cref{fig:density}a, while the $J_2$ model shows a fundamentally
    different behavior. The crack density is higher in the plastic
    case because of the additional tensile stresses caused at the edge
    of contacts by residual plastic
    deformations.}\label{fig:normal_density_plastic}
\end{figure}

Finally, \cref{fig:normal_density_plastic} shows the data of
\cref{fig:density_unnormalized}b normalized to compare the results to
\cref{eqn:p_crack}. While the saturation model seems to follow our GW
prediction (which is based on elasticity assumption), it is clear that
the $J_2$ model does not conform to our scaling predictions for
$P_\text{crack}$. However, in light of \cref{fig:plastic_sigma_ic}, it
is interesting to note that although $\sigma\ic$ is lowest for the
$J_2$ approach (indicating less interactions between contacts), the
latter has the largest crack nucleation likelihood, because of the
local effect of plastic residual deformations. As shown in
\cref{fig:tensile_single}, this local effect of plastic deformations
is stronger the more ductile a material is, as expected from
experimental data which shows that softer materials wear more.

\section*{Conclusion} We have investigated in this work the nucleation
of cracks at an elastic-plastic rough contact interface. This was
motivated by the necessity for an accurate description of the process
of crack nucleation for adhesive wear that includes plasticity. By
comparing a classical $J_2$ plasticity model and a saturation
plasticity approach commonly used in tribology, we have concluded that
the latter can only qualitatively reproduce the true contact (in some
cases) area and fails to give satisfactory results on local
quantities. This can be seen in the crack nucleation likelihood, which
is much higher in the $J_2$ approach. This is caused by plastic
residual deformations which increase tensile stresses in the vicinity
of contacts. In this regard, the saturation model is not applicable to
study crack nucleation because it does not capture plastic
deformations. We show with a single asperity analysis that more
ductile materials can have larger surface tensile stresses and
nucleate more cracks at the interface.

We have also showed that elastic interactions play a role in the crack
nucleation likelihood of a single contact. They may increase the
latter through proximity of contacts, or decrease it in shearing by
elastic shielding. This was further supported by an analytical
approach based on a Greenwood--Williamson model modified to take
interactions into account.

Future improvements of this work could include adhesive contact at the interface
due to surface energy. Attractive forces usually produce large stresses at the
surface, which may also contribute to the development of plastic zones in the
vicinity of contact edges.

\paragraph{Acknowledgments} All authors acknowledge the insightful
discussions with T.\ Brink and E.\ Milanese, as well as the financial
support of the Swiss National Science Foundation (grant \#162569,
``Contact mechanics of rough surfaces'').

\paragraph{Supplementary data} All codes used in this work are
available on
Zenodo~\citep{frerot_tamaas_2019,frerot_supplementary_2019a}.


\FloatBarrier{}
\appendix
\section{Inter-contact stress computation}\label{app:sigma_ic}
We first compute $\overline{\sigma_r}(r)$, which is the average radial stress outside contacts:
\begin{align*}
  \overline{\sigma_r}(r)
  &= \frac{1}{P(z^* \geq h)}
    \int_h^\infty{\kappa\frac{sR}{r^2}{(z^*-h)}^\frac{3}{2}e^{-z^*}\,\mathrm{d}z^*}\\
  & = \kappa \frac{3\sqrt{\pi}}{4}\cdot\frac{sR}{r^2}.
\end{align*}
We then suppose for simplicity that all contacts have the same radial stress $\overline{\sigma_r}(r)$. Accordingly, their hoop stress is $\overline{\sigma_\theta}(r) = -\overline{\sigma_r}(r)$ \citep{johnson_contact_1985a}. We assume that all contacts are spatially uniformly distributed with density $\eta_c$, so that the characteristic distance between contacts is $d_c = 1/\sqrt{\eta_c}$. We divide the infinite surface into concentric rings of width $d_c$ and diameters $d_i \in \{d_c, 2d_c, 3d_c,\ldots\}$. Each ring can be reduced to a circle of contacts with linear density $\sqrt{\eta_c}$. We now wish to compute for the sum of all circles of diameter $d_{1,2,\ldots}$ the largest stress eigenvalue at the center. For a single contact positioned at an angle $\theta$ on a circle of diameter $d_i$, the stress state in Cartesian coordinates is:
\begin{align*}
  \tens \sigma = \overline{\sigma_r}(d_i / 2)\begin{pmatrix}
    \cos(2\theta) & -\sin(2\theta)\\
    -\sin(2\theta) & -\cos(2\theta)\end{pmatrix}.
\end{align*}
For the circle number $i$, the expected number of contacts is $\pi i d_c \sqrt{\eta_c} = i\cdot\pi$, but we simplify by assuming the expected number to be $n_i := 3i$. The total stress state at the center, summing all contacts per circle and all circles:
\begin{align*}
  \tens \sigma = \sum_{i=1}^\infty{\sigma_r^i\sum_{k=1}^{n_i}{\begin{pmatrix}
        \cos(2\theta_k) & -\sin(2\theta_k)\\
        -\sin(2\theta_k) & -\cos(2\theta_k)\end{pmatrix}}},
\end{align*}
with $\sigma_r^i := \overline{\sigma_r}(id_c/2)$. The largest eigenvalue of $\tens \sigma$ is given by:
\begin{align*}
  {\lambda}^2& = {\left(\sum_{i=1}^\infty{\sigma_r^i\sum_{k=1}^{n_i}{\cos(2\theta_k)}}\right)}^2
             + {\left(\sum_{i=1}^\infty{\sigma_r^i\sum_{k=1}^{n_i}{\sin(2\theta_k)}}\right)}^2\\
  & = \sum_{i=1}^\infty{{(\sigma_r^i)}^2(c_i^2 + s_i^2)} + \sum_{i<j}^\infty{\sigma_r^i\sigma_r^j(c_i c_j + s_i s_j)},
\end{align*}
where:
\begin{align*}
  c_i := \sum_{k=1}^{n_i}\cos(2\theta_k),\quad s_i := \sum_{k=1}^{n_i}\sin(2\theta_k)
\end{align*}
The angular position of each contact is assumed to be uniformly distributed in $[0, 2\pi]$. The expected value of $\lambda$, which gives the inter-contact stress, is given by:
\begin{align*}
  \sigma\ic & = \mathbb{E}[\lambda] \leq \sqrt{\mathbb E[\lambda^2]} =   \sqrt{\sum_{i=1}^\infty{{(\sigma_r^i)}^2\mathbb E[c_i^2 \shp s_i^2]} \shp \sum_{i<j}^\infty{\sigma_r^i \sigma_r^j \mathbb E[c_i c_j \shp s_i s_j]}}.
\end{align*}
where we have used Jensen's inequality for a simple estimation. We note that:
\begin{align*}
  c_i^2 + s_i^2 & = {\left(\sum_{k=1}^{n_i}{\cos(2\theta_k)}\right)}^2 +
             {\left(\sum_{k=1}^{n_i}{\sin(2\theta_k)}\right)}^2\\
           & = \sum_{k=1}^{n_i}{\cos^2(2\theta_k)} + 2\sum_{k < l}^{n_i}{\cos(2\theta_k)\cos(2\theta_l)} + \sum_{k=1}^{n_i}{\sin^2(2\theta_k)} + 2\sum_{k < l}^{n_i}{\sin(2\theta_k)\sin(2\theta_l)}\\
                & = n_i + 2\sum_{k < l}^{n_i}{\cos(2(\theta_k - \theta_l))}\\
  c_i c_j + s_i s_j & = \sum_{k=1}^{n_i}{\cos(2\theta_k)}\sum_{l=1}^{n_j}{\cos(2\theta_l)} + \sum_{k=1}^{n_i}{\sin(2\theta_k)}\sum_{l=1}^{n_j}{\sin(2\theta_l)}\\
  & = \sum_{k=1}^{n_j}\sum_{l=1}^{n_j}\cos(2(\theta_k - \theta_l))
\end{align*}
Computing the expected value of the above expressions gives integrals of the form:
\begin{align*}
  \int_0^{2\pi}\int_0^{2\pi}\cos(2(\theta - \gamma))\,\mathrm{d}\theta\mathrm{d}\gamma = 0,
\end{align*}
and we simply obtain $\mathbb{E}[c_i^2 + s_i^2] = n_i = 3\cdot i$ and $\mathbb E[c_i c_j + s_i s_j] = 0$. Therefore:
\begin{align*}
  \sigma\ic & \leq \sqrt{\sum_{i=1}^\infty{{(\sigma_r^i)}^2n_i}}\\
            & \leq 3\kappa \sqrt{\pi}\frac{sR}{d_c^2} \sqrt{\sum_{i=1}^\infty{\frac{3}{i^3}}}\\
            & \leq 3\kappa\sqrt{\pi}sR\eta_c \sqrt{3\zeta(3)}
\end{align*}
where $\zeta$ is the Riemann Zeta function. We can now use the GW contact model to replace $sR\eta_c = sR\eta e^{-h} = A_c / (\pi A_0)$:
\begin{align*}
  \sigma\ic \leq 3\kappa\sqrt{\frac{3\zeta(3)}{\pi}}\cdot\frac{A_c}{A_0}.
\end{align*}
Note that only the $\sqrt{3\zeta(3)}$ term depends on the estimation from Jensen's inequality, so $\sigma\ic$ is indeed linear with respect to the contact ratio.

\section{Nucleation probability}\label{app:p_crack}
$P_\mathrm{crack}$ as defined in \cref{eqn:p_crack} is a conditional probability. It expresses the question ``knowing an asperity is in contact, what is the probability that a crack nucleates at the contact edge?'' The final expression for this probability is obtained by manipulating the inequality. We evaluate $\sigma_r$ at the contact radius $a = R\cdot s\sqrt{z^*-h}$:
\begin{align*}
  \sigma\I^\text{Hertz} = \sigma_r(R\cdot s\sqrt{z^*-h}) = \kappa{(z^*-h)}^\frac{1}{2},
\end{align*}
which is replaced in the inequality:
\begin{align*}
  \frac{\sigma_c}{\sigma\I^\text{Hertz} + \sigma\ic} & \leq \omega(\tau_j)\\
  \Leftrightarrow \frac{\sigma\I^\text{Hertz} + \sigma\ic}{\sigma_c} & \geq \frac{1}{\omega(\tau_j)}\\
  \Leftrightarrow \sigma\I^\text{Hertz} & \geq \frac{\sigma_c}{\omega(\tau_j)} - \sigma\ic\\
  \Leftrightarrow z^* & \geq {\left(\frac{\sigma_c/\omega(\tau_j) - \sigma\ic}{\kappa}\right)}^2 + h.
\end{align*}
Let us call $X$ the event corresponding to the above inequality. We have:
\begin{align*}
  P_\text{crack} & = P(X\ \big|\ z^* \geq h) \\
                 & = \frac{P(X\ \mathbf{and}\ z^* \geq h)}{P(z^* \geq h)}\\
                 & = \frac{P(X)}{P(z^* \geq h)}\\
                 & = \exp\left(-{\left(\frac{\sigma_c/\omega(\tau_j) - \sigma\ic}{\kappa}\right)}^2\right),
\end{align*}
since $z^*$ follows the canonical exponential distribution.

\bibliographystyle{elsarticle-harv}
\bibliography{biblio}

\end{document}